\documentclass[12pt,letterpaper]{article}                  
\usepackage{osajnl}
\usepackage{overcite,hyperref}
\usepackage[latin1]{inputenc}

\begin{document}

\title{Conditions for waveguide decoupling in square-lattice photonic crystals}

\author{T. Koponen$^1$, A. Huttunen$^2$ and P. Törmä$^1$}

\affiliation{$^1$Department of Physics, University of Jyväskylä\\PB 35
  (YFL), FIN-40014 Jyväskylä, Finland\\$^2$Laboratory of Computational
  Engineering, Helsinki University of Technology,\\P.O.Box 9203,
  FIN-02015, Finland}

\begin{abstract}
We study coupling and decoupling of parallel waveguides 
in two-dimensional square-lattice photonic crystals.
We show that the waveguide coupling is prohibited 
at some wavelengths when there is an odd number of rows between 
the waveguides. In contrast, decoupling does not take place when there is even number of 
rows between the waveguides. Decoupling can be used to avoid cross talk between
adjacent waveguides.
\end{abstract}


\maketitle 

Two-dimensional photonic crystals are promising candidates for 
implementing integrated optical components \cite{Joannopoulos}. 
Optical waveguiding in two-dimensional photonic crystals is achieved 
by introducing line defects in the structure that is otherwise 
periodic in two dimensions \cite{Joannopoulos_book}. 
Two parallel waveguides can be used as a directional waveguide coupler 
\cite{Koshiba, Tokushima, Boscolo, Martinez, Qiu}.
On the other hand, it might be desirable to decouple the two waveguides to 
minimize cross talk between them, for instance when envisioning
closely packed photonic wires in integrated optical circuits.\cite{Kuchinsky}

We study the coupling between two parallel waveguides 
in a square-lattice photonic crystal
and find that the decoupling of the waveguides depends on the number 
of rods between the waveguides. 
If there is an odd number of rods between the waveguides, 
they are decoupled at a defined wavelength.
In case of an even number of rods, the waveguides 
are coupled at all wavelengths.
Previous studies such as \cite{Boscolo} have considered an even number
of rods and therefore not demonstrated the decoupling behavior.

The studied geometry is a two-dimensional photonic crystal of
cylindrical dielectric rods in a square-lattice in air. 
The dielectric
constant of the rods is taken to be \(\epsilon = 8.9\epsilon_0\), their radius \(r = 0.2a\) and
the lattice constant \(a = 512\,\)nm. The photonic crystal has
a large TE (electric field aligned with the cylinders) band gap around
\(\omega = 0.8\pi c/a\), where \(a\) is the 
lattice constant of the crystal. With \(a = 512\,\)nm, this gap is in
the wavelength range from \(1100\,\)nm to \(1600\,\)nm.
Two parallel
waveguides are formed in the structure by 
removing two parallel rows of rods. The number of rods
between the two waveguides is varied. In Fig. 1 (a) and (c) we show two examples
of the geometries, i.e. one and two rows of rods between the waveguides. We
have considered 1-7 rows of rods between the waveguides.

\begin{figure}
\centering
\includegraphics{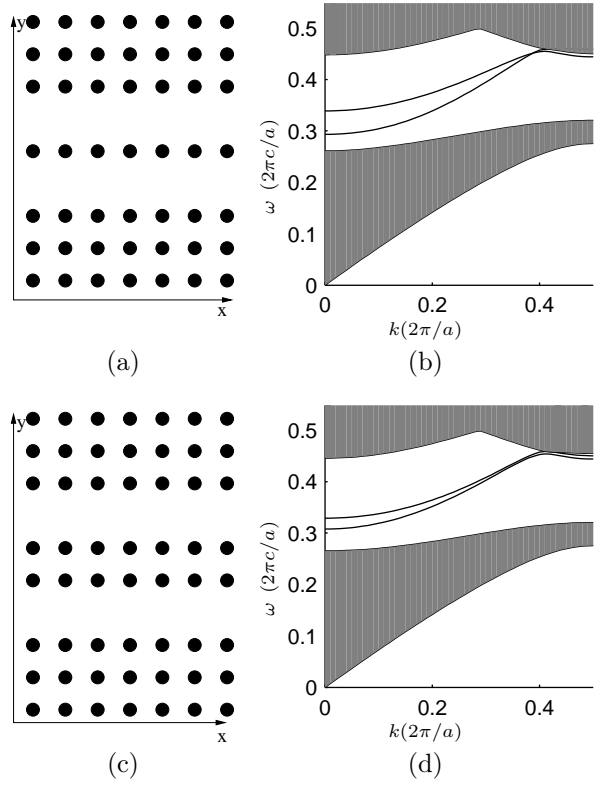}
\caption{Geometries and band structures for one [(a) and (b)] and two 
[(c) and (d)] rows of dielectric rods between two parallel waveguides. 
The \(z\)-direction points out of the plane.}
\end{figure}

The guided eigenmodes supported
by this geometry have two possible parities with
respect to the symmetry axis between the waveguides. These modes can
be classified by the parity of the \(z\)-component of the electric
field, \(E_z\). Following directly from Maxwell's equations, 
the parity of \(H_x\) is always opposite to the parity of \(E_z\), and
the parity of \(H_y\) is always the same as that of
\(E_z\). According to the parity of \(E_z\), the two eigenmodes can be
labeled ``even'' and ``odd''. 
Here \(z\)
is the direction of the cylinders (out of plane), \(x\) is along the
waveguides and \(y\) is orthogonal to the cylinders and the waveguides.

When the system only supports two guided modes, any signal
\(\Psi(x,y,t)\) with a definite frequency \(\omega\) propagating in
the system can be written as a superposition of these two eigenmodes
\begin{equation}
\Psi(x,y,t) = \psi_E(x,y,t)\exp(ik_Ex) + \psi_O(x,y,t)\exp(ik_Ox).
\end{equation}
Here, \(\psi_E\) and \(\psi_O\) stand for even and odd eigenmodes and
\(k_E\) and \(k_O\) for the corresponding values of \(k\). The
spatial dependence of \(\psi_E\) and \(\psi_O\) is lattice-periodic.
This kind of a superposition gives rise to beating between the eigenmodes. 
The plane wave terms in Eq. (1) are in the same phase when \(x=0\)
and in the opposite phase when \(x = \pi/|k_O-k_E|\). The beating
wavelength is therefore
\begin{equation}
\kappa = \frac{2\pi}{|k_O - k_E|}.
\end{equation}
When the eigenmodes are in the same phase, their superposition has
most of its energy in one of the waveguides and when in opposite phase, 
in the other. The propagating signal
oscillates between the two waveguides with the characteristic
wavelength \(\kappa\) given above.
This is the mechanism applied e.g. in waveguide couplers. Note that
coupling can be realized also by defects between the waveguides \cite{Fan} 
or coupling can be between a waveguide and a defect \cite{Noda}.

The beating wavelength \(\kappa\) becomes infinite when \(k_E=k_O\).
This means that there is no energy
transfer between the waveguides, i.e. the waveguides are
decoupled. For the \(k\)-values to be identical, the bands of the even 
and odd eigenmodes have to cross. If they avoid crossing, \(\kappa\) 
is always finite and the two waveguides cannot be decoupled.

We have calculated the band structures of two parallel waveguides in 
a square-lattice photonic crystal
with the MIT Photonic Bands\cite{MPB} program. The band
structures for the case of one and two rods between the waveguides are
shown in Fig. 1 (b) and (d), respectively. 
It can be seen that the bands for the even and odd
eigenmodes cross in Fig. 1 (b), but do not cross in Fig. 1 (d). 
We calculated the band structures for 1-7 rows between the waveguides and 
found that, for geometries with an odd number rods between the
waveguides, the bands for odd and even eigenmodes cross, whereas they
never cross when there is an even number of rods between the
waveguides. 

We calculated the coupling wavelengths \(\kappa\) [Eq. (2)] from the band
structures (Fig. 1) and also with the Finite-Difference Time-Domain
Method\cite{Taflove}. Results from both methods are shown in Fig. 2 for the same geometries 
as considered in Fig. 1.
There is a singularity, corresponding to decoupling, in case there is an odd number of rows 
between the waveguides [see Fig. 2 (a)].
In Fig. 2 (b) the value of $\kappa$ is always finite. We found such
behavior for 1-7 rows of rods between the waveguides. We performed the
same calculations with FDTD for 1-4 rows between the waveguides.
The calculations using the band structures and FDTD simulations 
are in excellent agreement.\cite{comment}

\begin{figure}
\centering
\includegraphics{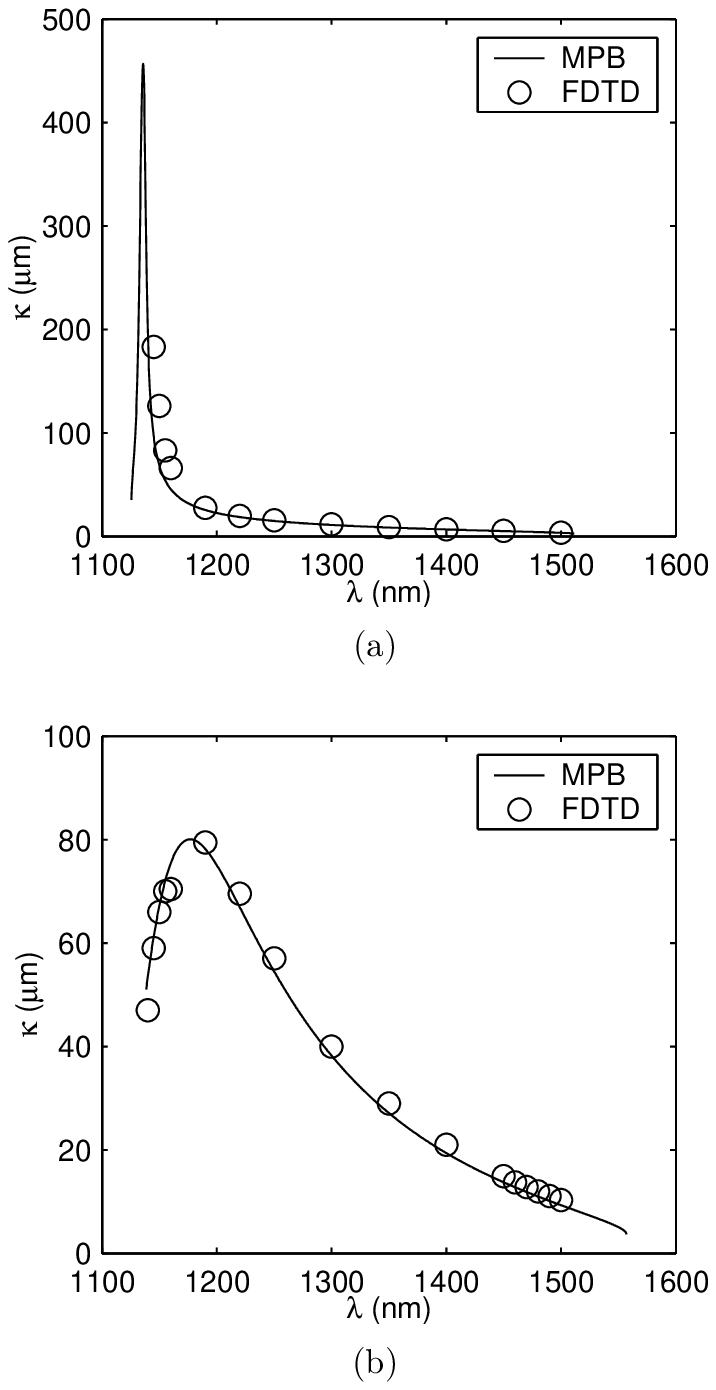}
\caption{The coupling distance as a function of the wavelength of the
  light propagating in the waveguides,
  for the geometries with one (a) and two (b) rows of rods between the waveguides.
The solid curve is calculated using the MIT Photonic Bands program and
the circles are calculated by FDTD.}
\end{figure}

In order to explain the strong effect of the geometry one has to consider the field 
distributions of the eigenmodes.
The $E_z$ and $H_y$ components of the odd eigenmode have a node 
in the symmetry plane of the structure. 
The parity of the $H_x$ component is the opposite of the parities 
of the $E_z$ and $H_y$ components.
When there is an odd number of rods between the waveguides, the nodes 
of the $E_z$ and $H_y$ components of the odd eigenmode are 
in the center of a dielectric rod. 
For the even eigenmode, the $E_z$ and $H_y$ components are nonzero at
the symmetry plane.
It is known that the more the fields are inside the material of 
high dielectric constant, the smaller the energy.
Thus at small values of the wave vector, the energy of the even eigenmode is 
smaller than that of the odd eigenmode.
The bands cross at some value of the wave vector.
This is because the relative power of the $H_y$ component compared to the
power of the $H_x$ component increases with increasing values of 
the wave vector. Then $H_y$ starts to determine the effective parity of
the mode.
Thus at large values of the wave vector the effective parities of the
eigenmodes change and thus the bands cross. 
When there is an even number of rods between the waveguides, the node
of the odd eigenmode is in air and the effective parity does not have 
such an effect to the energies of the eigenmodes. In this case the odd 
eigenmode has a lower energy at all values of the wave vector. This
explanation corresponds to the behavior of the eigenmodes in the
particular geometry considered in this paper. In general, our findings
demonstrate that symmetry properties of a photonic crystal waveguide
pair, especially parity effects, can be used to design the waveguide
properties, for instance, to produce complete decoupling. In this
sense photonic crystal waveguides posses an additional degree of
freedom compared to traditional dielectric waveguides.

\end{document}